\documentclass[letter]{aa}
\usepackage{graphicx}
\usepackage{txfonts}
\usepackage{natbib}
\bibpunct{(}{)}{;}{a}{}{,} 
\begin{document}
\title{Fundamental properties and atmospheric structure of the red 
supergiant VY~CMa based on VLTI/AMBER spectro-interferometry
\thanks{Based on observations made with the VLT Interferometer (VLTI)
at Paranal Observatory under programme ID 386.D-0012}}
\author{
M.~Wittkowski\inst{1}\and
P.~H.~Hauschildt\inst{2}\and
B.~Arroyo-Torres\inst{3}\and
J.~M.~Marcaide\inst{3}
}
\institute{
ESO, Karl-Schwarzschild-Str. 2,
85748 Garching bei M\"unchen, Germany,
\email{mwittkow@eso.org}
\and
Hamburger Sternwarte, Gojenbergsweg 112, 21029, Hamburg, Germany
\and
Dpt. Astronomia i Astrof\'isica, Universitat de Val\`encia, 
C/ Dr. Moliner 50, 46100, Burjassot, Spain
}
\date{Received \dots; accepted \dots}
\abstract{}
{We investigate the atmospheric structure and fundamental 
properties of the red supergiant VY~CMa.}
{We obtained near-infrared spectro-interferometric observations
of VY~CMa with spectral resolutions of 35 and 1500 
using the AMBER instrument at the VLTI.}
{The visibility data indicate the presence of molecular
layers of water vapor and CO in the extended atmosphere with an
asymmetric morphology. The uniform disk diameter in the 
water band around 2.0\,$\mu$m is increased by $\sim$20\% compared
to the near-continuum bandpass at 2.20--2.25\,$\mu$m, and in the CO
band at 2.3--2.5\,$\mu$m it is increased by up to $\sim$50\%.
The closure phases indicate relatively small deviations from point 
symmetry close to the photospheric layer, and stronger 
deviations in the extended H$_2$O and CO layers.
Making use of the high spatial and spectral
resolution, a near-continuum bandpass can be isolated from
contamination by molecular and dusty layers, and the 
Rosseland-mean photospheric angular diameter
is estimated to 11.3$\pm$0.3\,mas based on a PHOENIX atmosphere model.
Together with recent high-precision estimates
of the distance and spectro-photometry, this estimate
corresponds to a radius of 1420$\pm$120\,$R_\odot$ and an 
effective temperature of 3490$\pm$90\,K.
}
{VY~CMa exhibits asymmetric, possibly clumpy, atmospheric
layers of H$_2$O and CO, which are not co-spatial, within a larger 
elongated dusty envelope. Our revised
fundamental parameters put VY~CMa close to the Hayashi limit of recent 
evolutionary tracks of initial mass 25\,$M_\odot$ with rotation
or 32\,$M_\odot$ without rotation, shortly before evolving blueward
in the HR-diagram.}
\titlerunning{Fundamental properties of VY CMa}
\maketitle
\section{Introduction}
VY~Canis~Majoris (VY~CMa) is one of the most luminous and most 
massive red supergiants, and thus most likely progenitors
of a core-collapse supernova (SN) in our Galaxy. Because of its 
brightness and importance, it has traditionally been extensively 
observed at many wavelengths. Currently, VY~CMa experiences a renewed
and increased interest because of observations at new facilities 
including, for instance, the Herschel space observatory 
\citep{royer10} and the submillimeter array \citep{fu11}, 
as well as because of new results regarding the complexity of its 
molecular envelope \citep[e.g.,][]{ziury07}.

While is has been clear that VY CMa's fundamental parameters are extreme, 
their exact values have been debated during the last decade with
luminosities between $6\times10^4 L_\odot$ and $5\times10^5 L_\odot$,
effective temperatures between 2700\,K and 3650K, radii between
600\,$R_\odot$ and 3000\,$R_\odot$, and initial masses between 12\,$M_\odot$
and 40\,$M_\odot$ \citep{smith01,monnier00,monnier04,massey06}. 
Very recently, precise parallax measurements by \citet{choi08}
and \citet{zhang12} significantly improved the 
distance estimate to about 1.2\,kpc, which together with recent accurate 
broad-band photometry by \citet{smith01} points to a luminosity of 
about $3\times10^5 L_\odot$. 

VY~CMa is 
embedded in a large optical nebula \citep{herbig72} 
and is also an extended object in the infrared \citep{mccarthy75}.
Several studies have investigated the asymmetric circumstellar environment
in detail and revealed a complex asymmetric morphology
\citep[e.g.][]{wittkowski98,monnier99,monnier00,monnier04,smith01,muller07,humphreys07,jones07}.
Recently, \citet{smith09} proposed a scenario of VY~CMa's circumstellar
environment that consists of fast and dense CO/dust cloudlets
that are ejected into a larger relatively
slow and less dense asymmetric envelope due to a strong and variable mass loss.
\section{Observations and data reduction} 
\label{sec:obs}
\begin{table}
\caption{Log of our VLTI/AMBER observations}
\label{tab:vlti}
\centering
\begin{tabular}{lrrrrrrrrr}
\hline\hline
Target   & Date       & Time  & Mode    & Baseline        & PA     \\ 
         & 2011-03    & UT    &         & m               & $\deg$ \\\hline
VY CMa   & -06 & 1:42  & Low-JHK & 15.8/31.6/47.5  & $-$75    \\
VY CMa   & -06 & 3:05  & MR-K 2.3& 14.4/28.7/43.0  & $-$83    \\
VY CMa   & -06 & 4:14  & Low-JHK & 11.8/23.6/35.4  & $-$90    \\
VY CMa   & -07 & 0:55  & Low-JHK & 16.0/32.0/48.0  & $-$70    \\
VY CMa   & -07 & 1:35  & Low-JHK & 15.9/31.7/47.6  & $-$75    \\
VY CMa   & -07 & 3:02  & Low-JHK & 14.3/28.6/42.9  & $-$83    \\
VY CMa   & -07 & 4:29  & MR-K 2.1& 11.2/22.3/33.5  & $-$89    \\\hline
\end{tabular}
\tablefoot{
The AMBER instrument modes are Low-JHK (range 1.2--2.5\,$\mu$m, 
spectral resolution $R\sim30$), MR-K 2.1 (1.92--2.26\,$\mu$m, $R\sim1500$), 
and MR-K 2.3 (2.12--2.47\,$\mu$m, $R\sim1500$).
The baseline denotes the projected baseline lengths, and the 
position angle (PA) the projected position angle on sky (east of north).}
\end{table}
We observed VY~CMa with the VLTI/AMBER instrument in visitor mode
on 6 and 7 March 2011.
Table~\ref{tab:vlti} lists the details of our observations.
We used the E0-G0-H0 configuration of the VLTI,
providing projected baseline lengths between 11.2\,m and 48.0\,m.
We obtained 5 observations using the low resolution mode with a 
spectral resolution of $R\sim30$ at different hour angles, thus
different projected baseline lengths and angles. We also obtained 
one observation
using each of the two $K$-band medium resolution modes, together covering 
wavelengths between 1.92\,$\mu$m and 2.47\,$\mu$m at a spectral resolution of
$R\sim1500$. All VY~CMa observations are embedded between observations
of an interferometric calibration star. We chose HR~3052 
(spectral type K5\,III, angular diameter 2.62$\pm$0.19\,mas) and 
1~Pup (K5\,III, 3.85$\pm$0.28\,mas) from the ESO Calibrator Selector
CalVin based on the catalog by \citet{lafrasse10}. 

We obtained raw visibility and closure phase using the {\tt amdlib} data 
reduction package \citep{tatulli07,chelli09}, version 3.0.3. 
We performed an absolute wavelength calibration 
as in \citet{wittkowski11}.
Calibrated visibility spectra were
obtained by using an average of two transfer function
measurements taken before and after each science observation.

Finally, we took into account that VY~CMa shows an extended
dust shell that exceeds the AMBER field of view (FoV).
In this case, the calibrated visibility values
are normalized to the flux within the AMBER FoV and not to the larger
flux of the more extended target, which increases the fraction of the 
correlated flux and thus the visibility values \citep[cf.][]{driebe09}. 
Here, this scaling factor was estimated using the results
by \citet{wittkowski08} based on bispectrum speckle interferometry with a 
sufficiently large FoV. They obtained a relative flux contribution
of an unresolved component of 9\%/50\% at wavelengths of 
1.28\,$\mu$m/2.17\,$\mu$m. With our longer baselines, their extended
component is fully resolved, and our visibility data describe their
unresolved component. Thus, we used a model of a central uniform disk (UD) and a 
fully resolved component, and scaled our visibility values such that 
we obtained the same relative flux contribution of our central component.
The resulting scale factor for the visibility modulus was 0.45/0.90 
in the $J$/$K$ bands, and we used an interpolated $H$-band value of 0.65.
We note that this procedure is applied in order to remove
the instrumental signature of the limited FoV, which may be useful
for a future combination of our data with data obtained
by other instruments. However, the choice of the scaling factor does 
not affect the results presented in the following sections, except for
the fraction of the flux that is attributed to the central 
stellar component, as the visibilities are fitted with a two-component
model where the flux ration is a free parameter (see Sect.~\ref{sec:fundpar}).
\section{Atmospheric structure and morphology}
\begin{figure}
\centering
\resizebox{1\hsize}{!}{\includegraphics{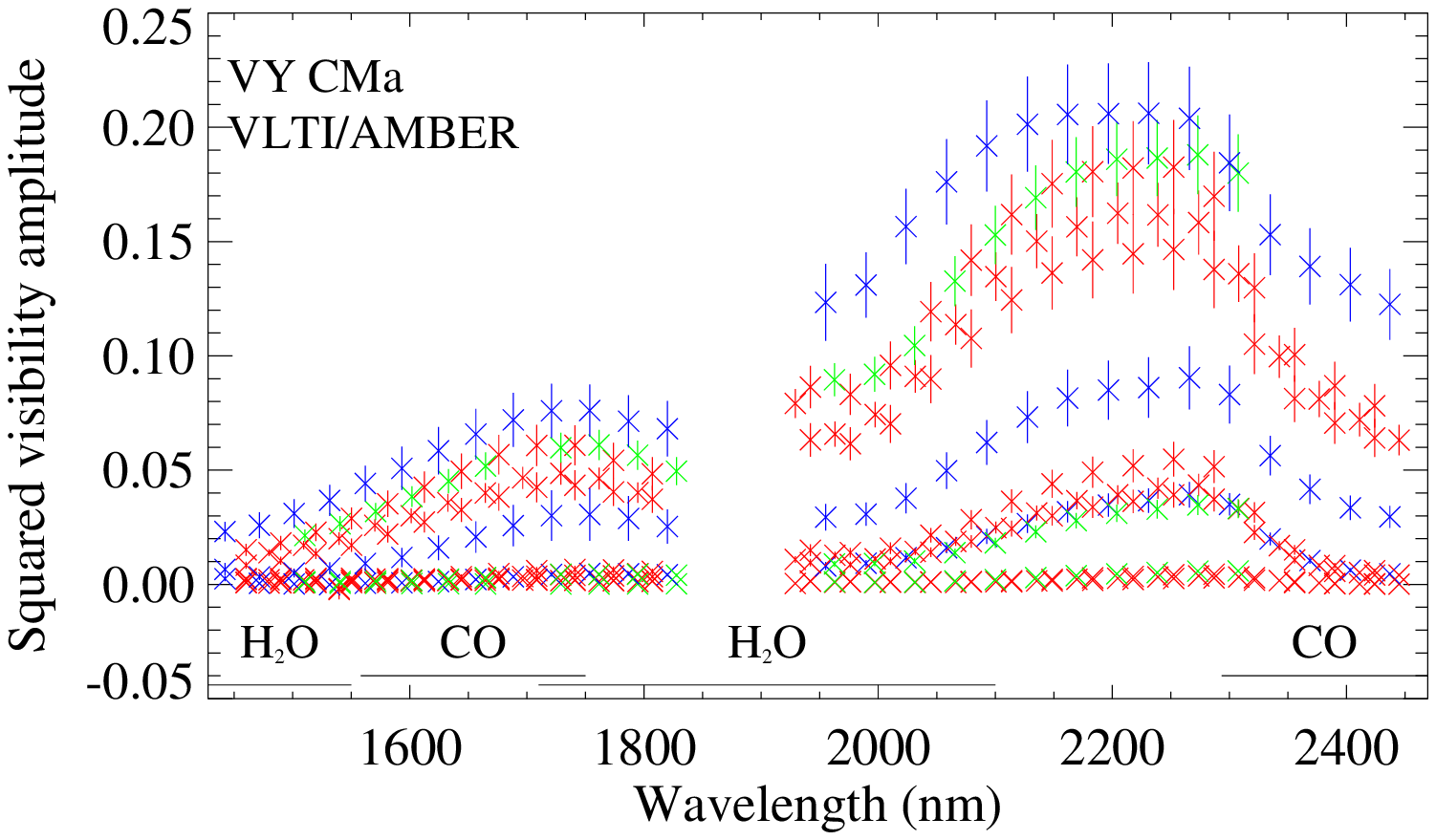}}
\resizebox{1\hsize}{!}{\includegraphics{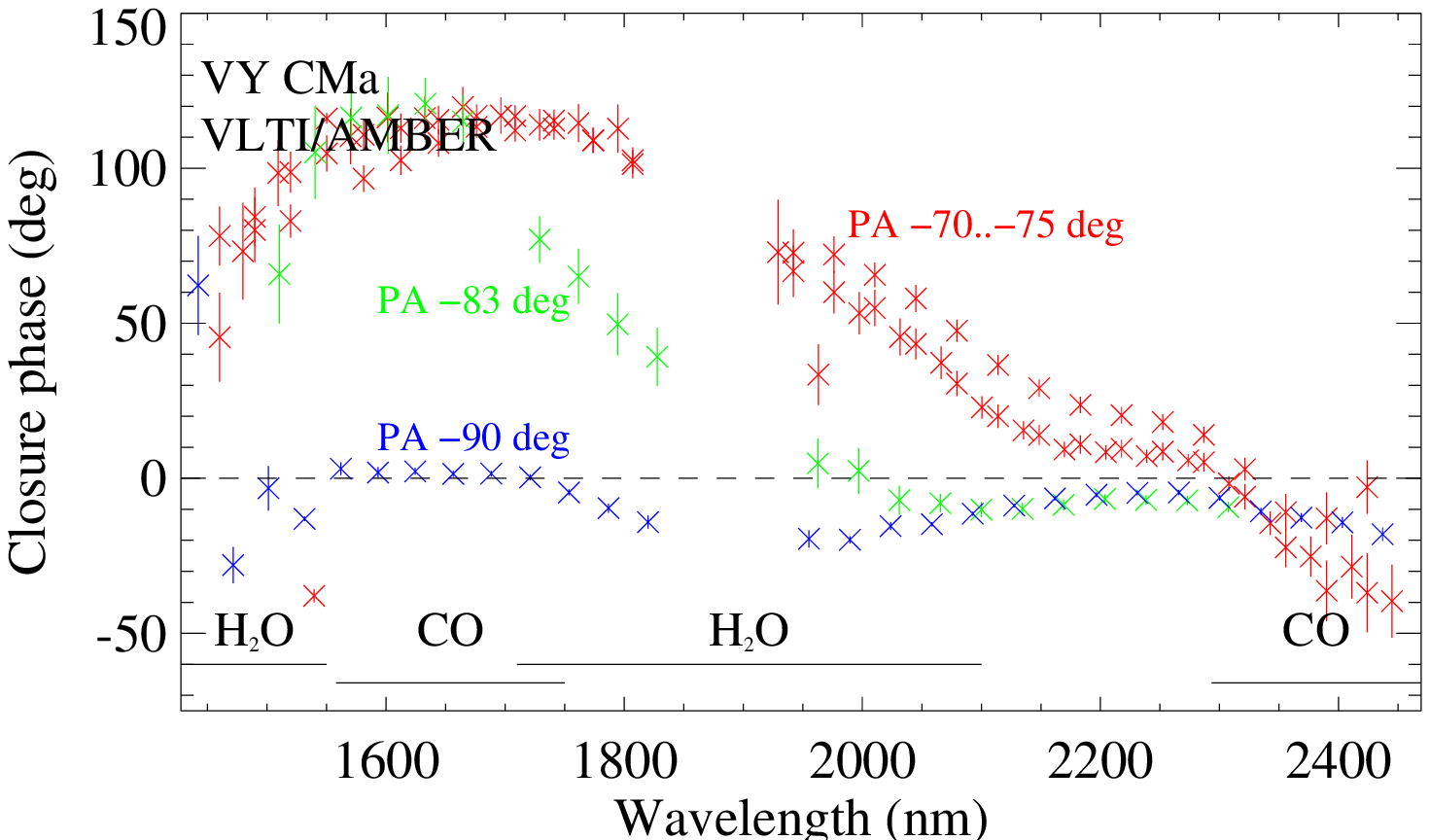}}
\caption{Squared visibility amplitudes (top) and closure phases (bottom)
of VY~CMa obtained with the low resolution mode of the AMBER instrument.
Also shown are the positions of CO and H$_2$O bands 
after \protect\citet{lancon00}. The different colors indicate
our different ranges of PA (cf. Table~\ref{tab:vlti}).}
\label{fig:lowres}
\end{figure}
\onlfig{2}{
\begin{figure}
\centering
\resizebox{1\hsize}{!}{\includegraphics{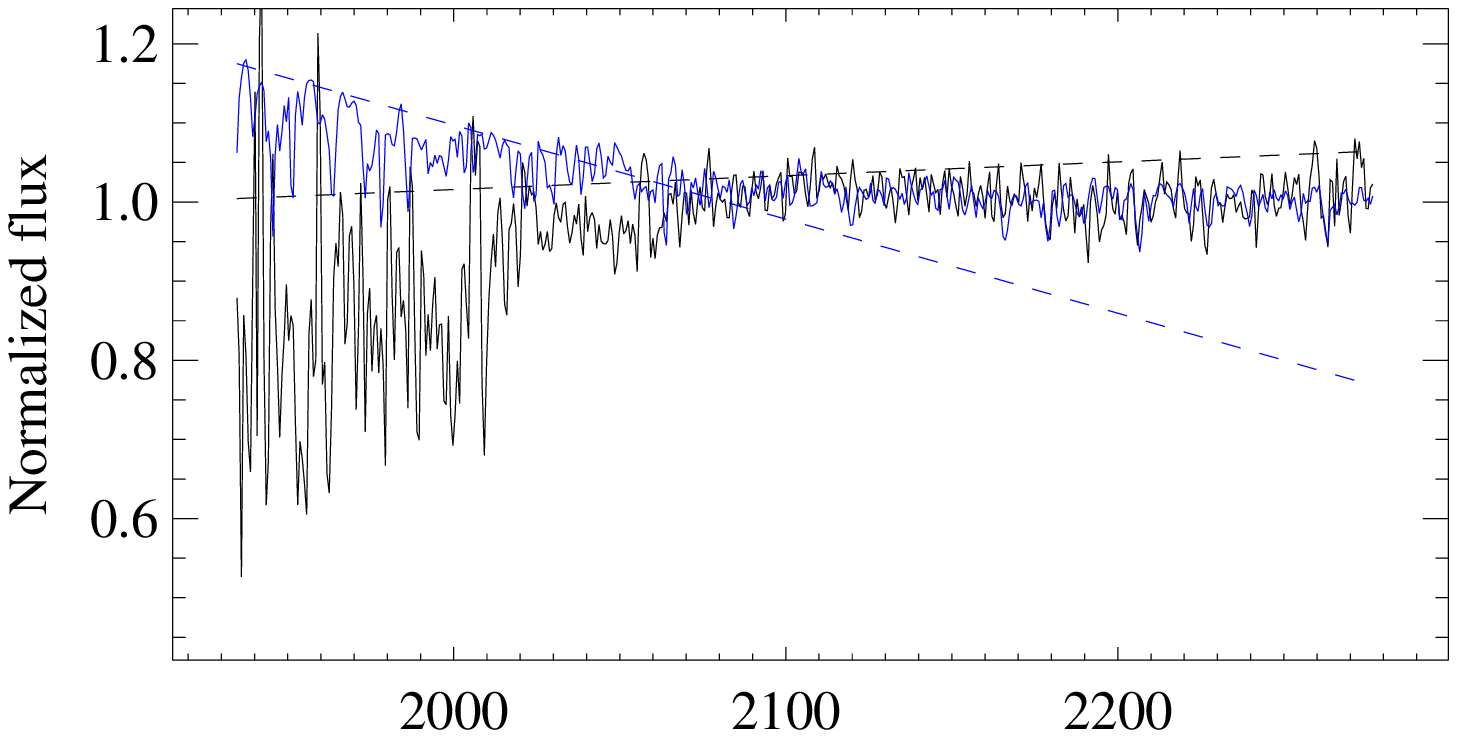}}
\resizebox{1\hsize}{!}{\includegraphics{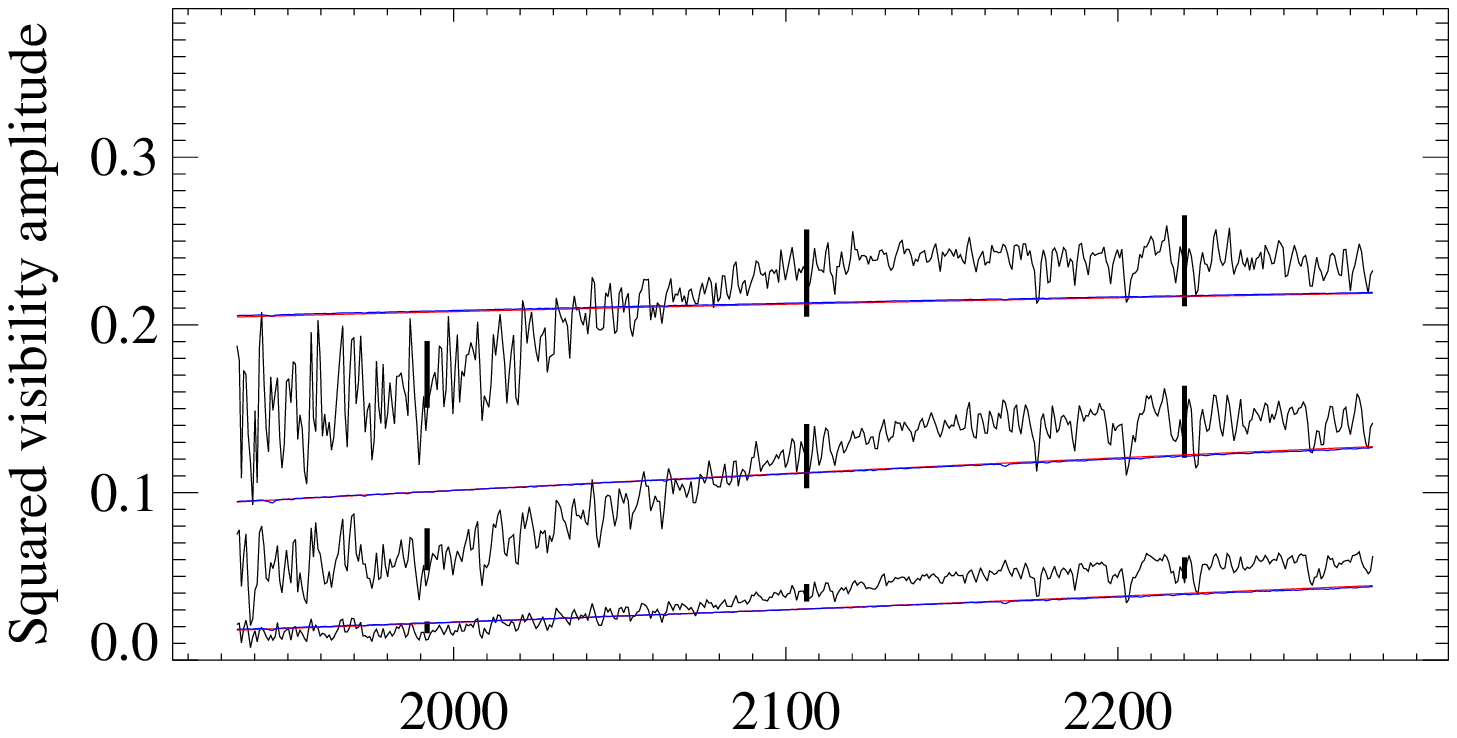}}
\resizebox{1\hsize}{!}{\includegraphics{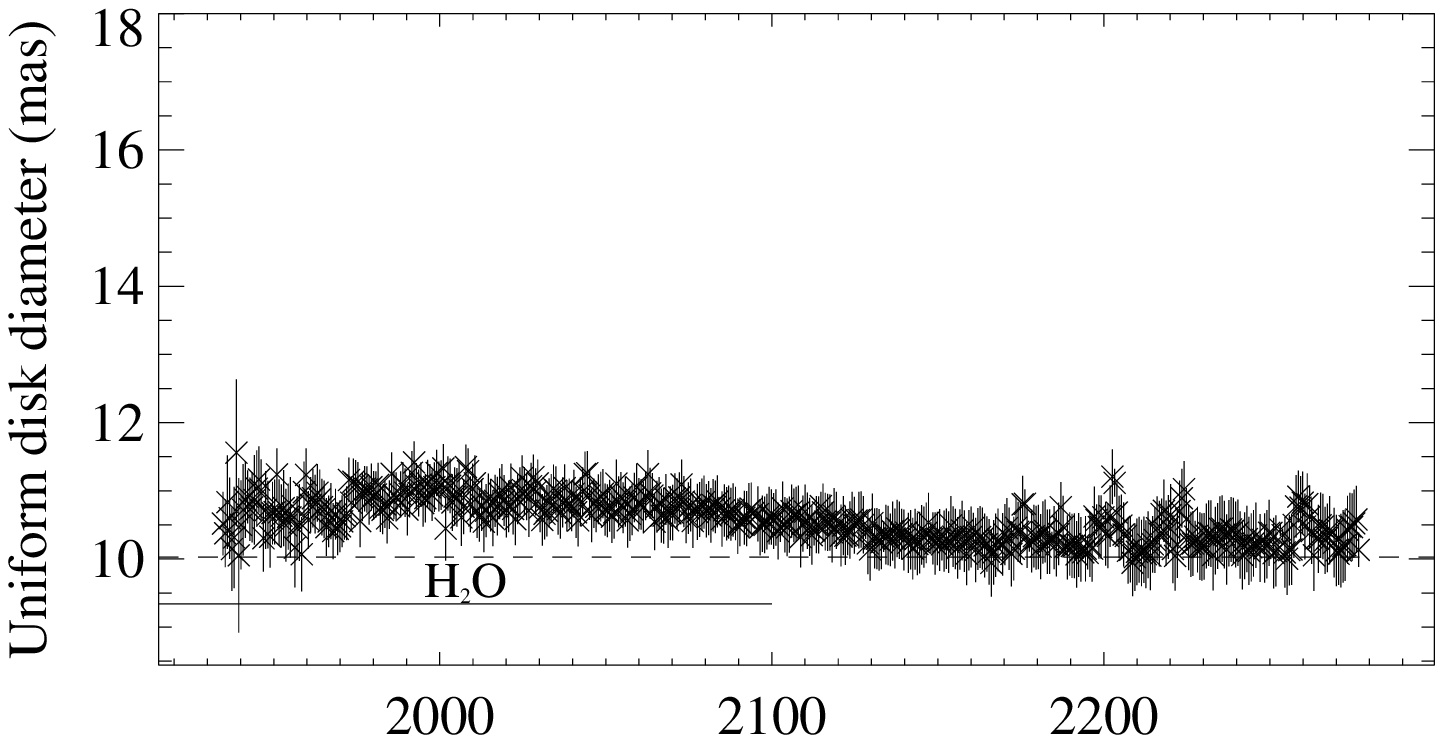}}
\resizebox{1\hsize}{!}{\includegraphics{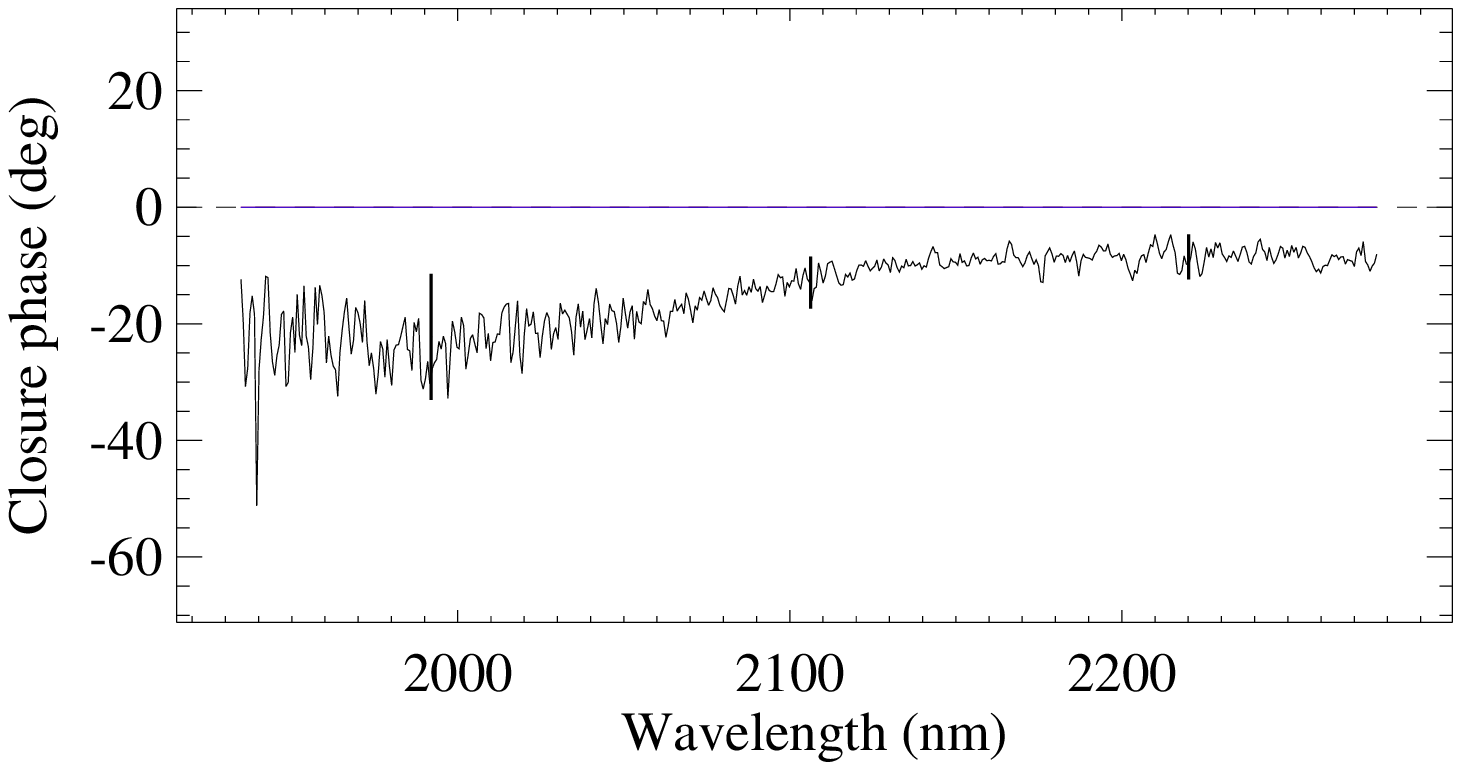}}
\caption{Normalized flux, squared visibility amplitudes, UD diameters,
and closure phases (from top to bottom) of VY~CMa obtained with the MR-K 2.1 $\mu$m
mode
(PA $-89\deg$). The top panels also show fits of a UD (red) and a 
{\tt PHOENIX}
atmosphere model (blue). In these model visibility fits, the stellar component
contributes $\sim$50\% of the total flux, and the remaining part
is attributed to an over-resolved dust shell. The observed and model
flux spectra show different slopes because of the additional dust shell
that is not present in the model, and we have subtracted a best-fit linear
function for a better comparison of the spectral features. The slopes that
have been compensated are indicated by the dashed lines.}
\label{fig:mr21}
\end{figure}
}
\onlfig{3}{
\begin{figure}
\centering
\resizebox{1\hsize}{!}{\includegraphics{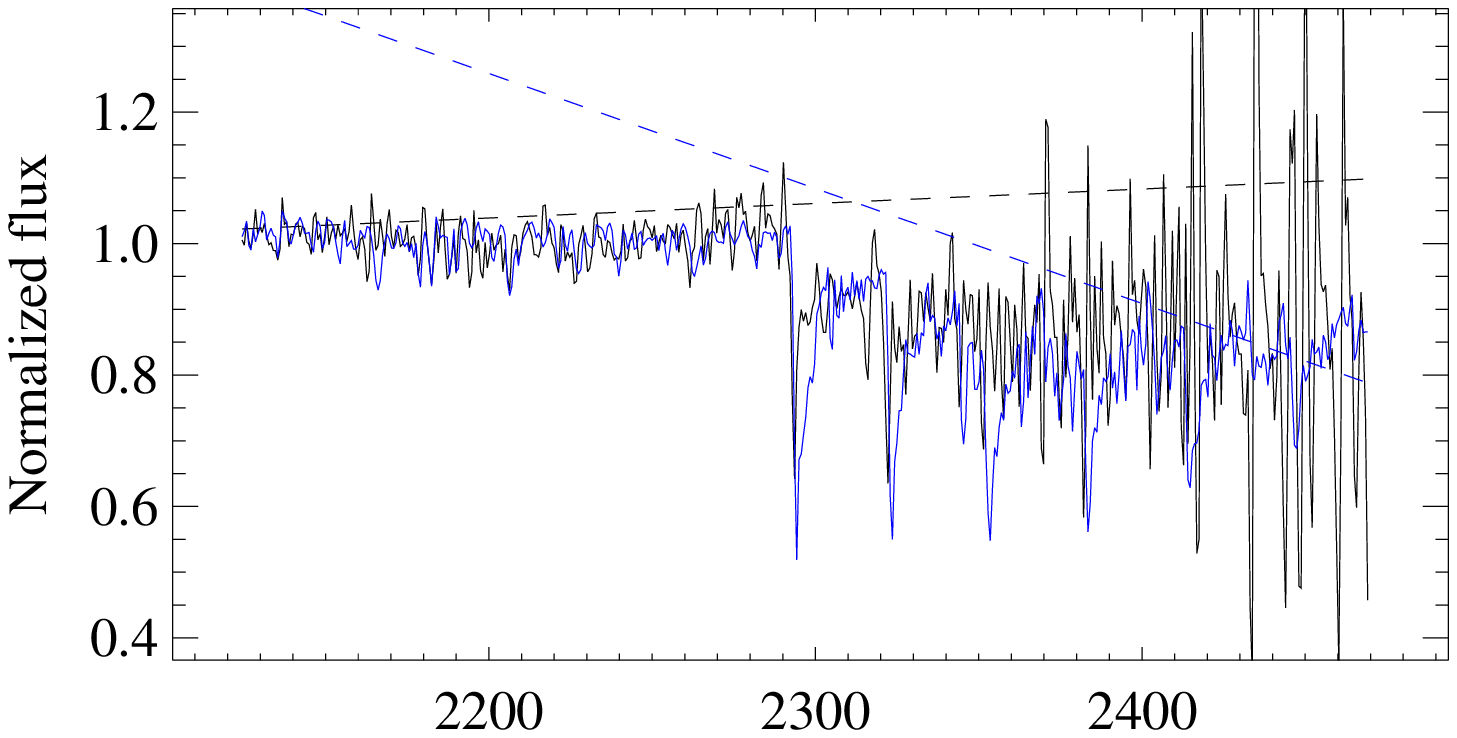}}
\resizebox{1\hsize}{!}{\includegraphics{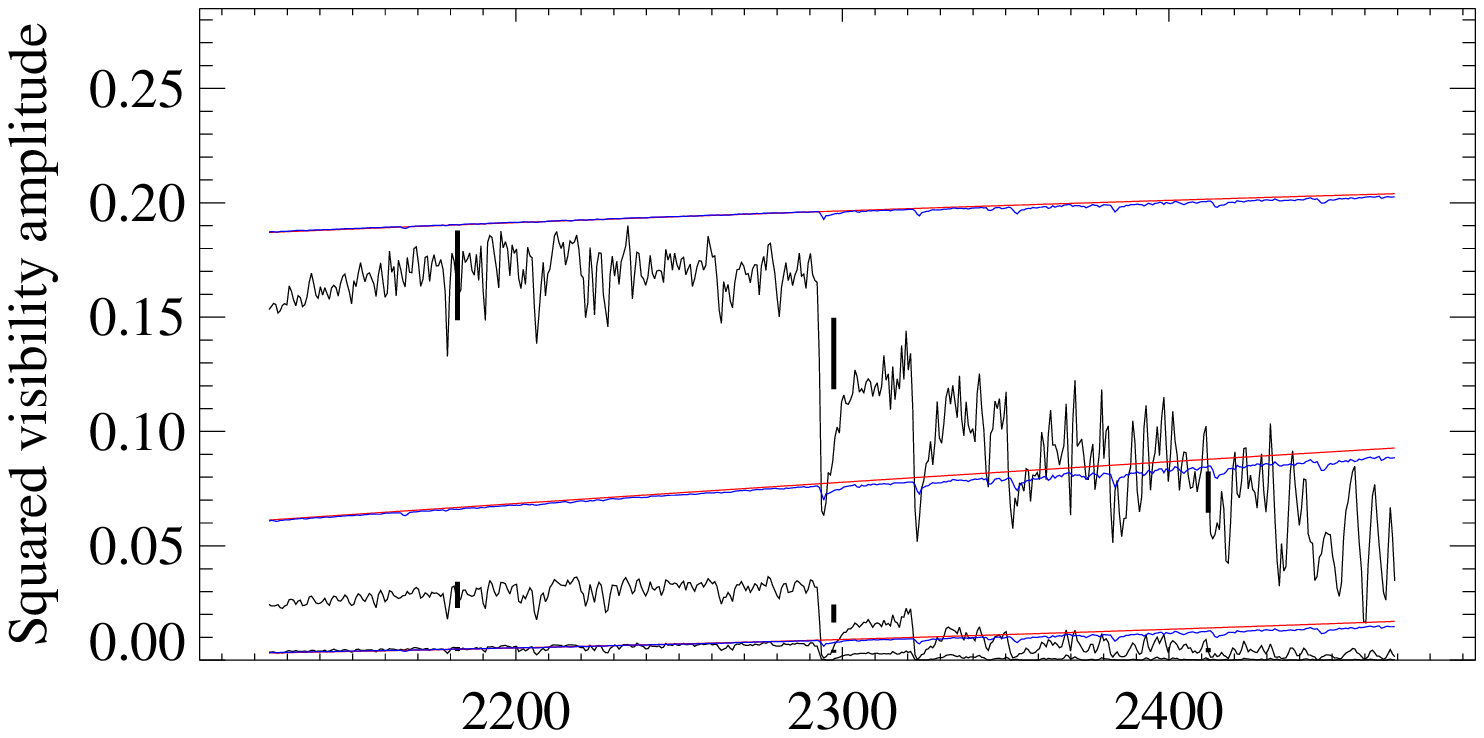}}
\resizebox{1\hsize}{!}{\includegraphics{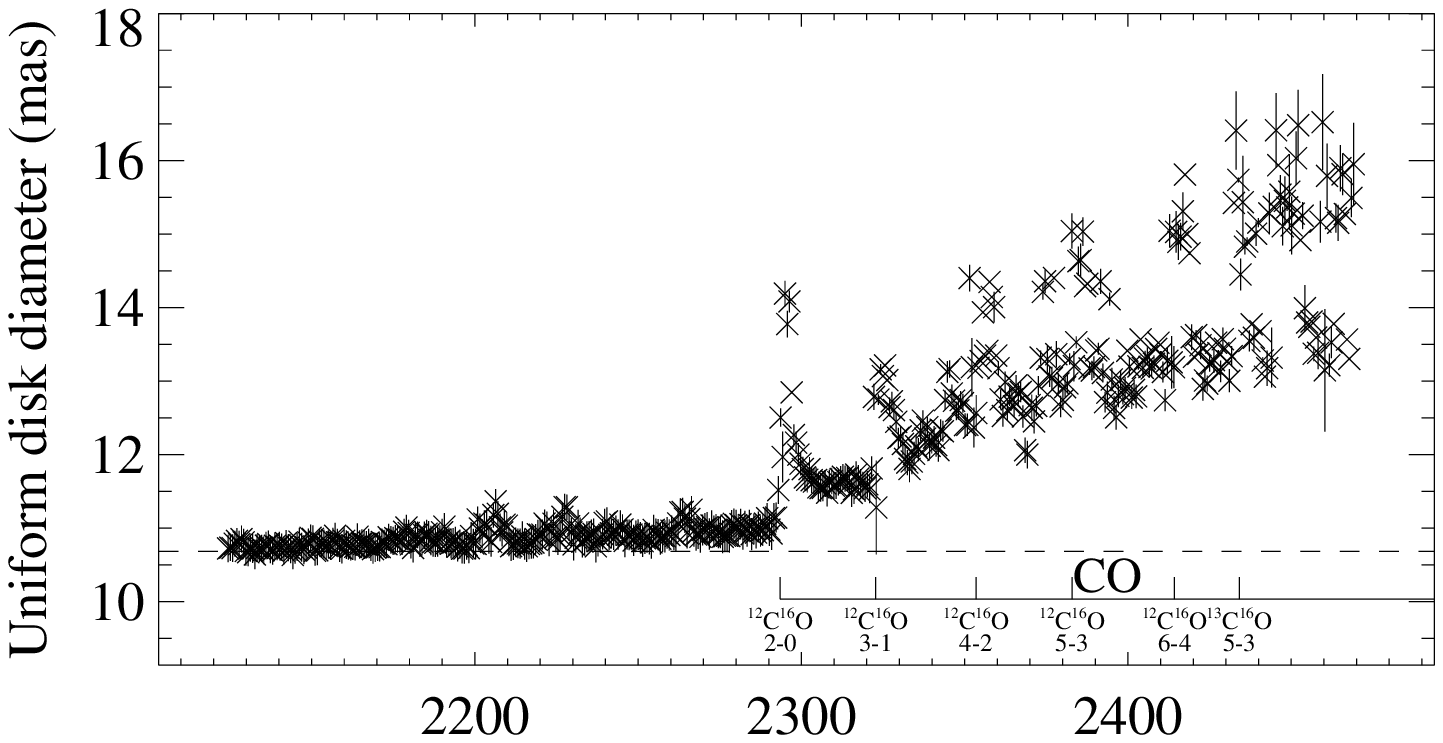}}
\resizebox{1\hsize}{!}{\includegraphics{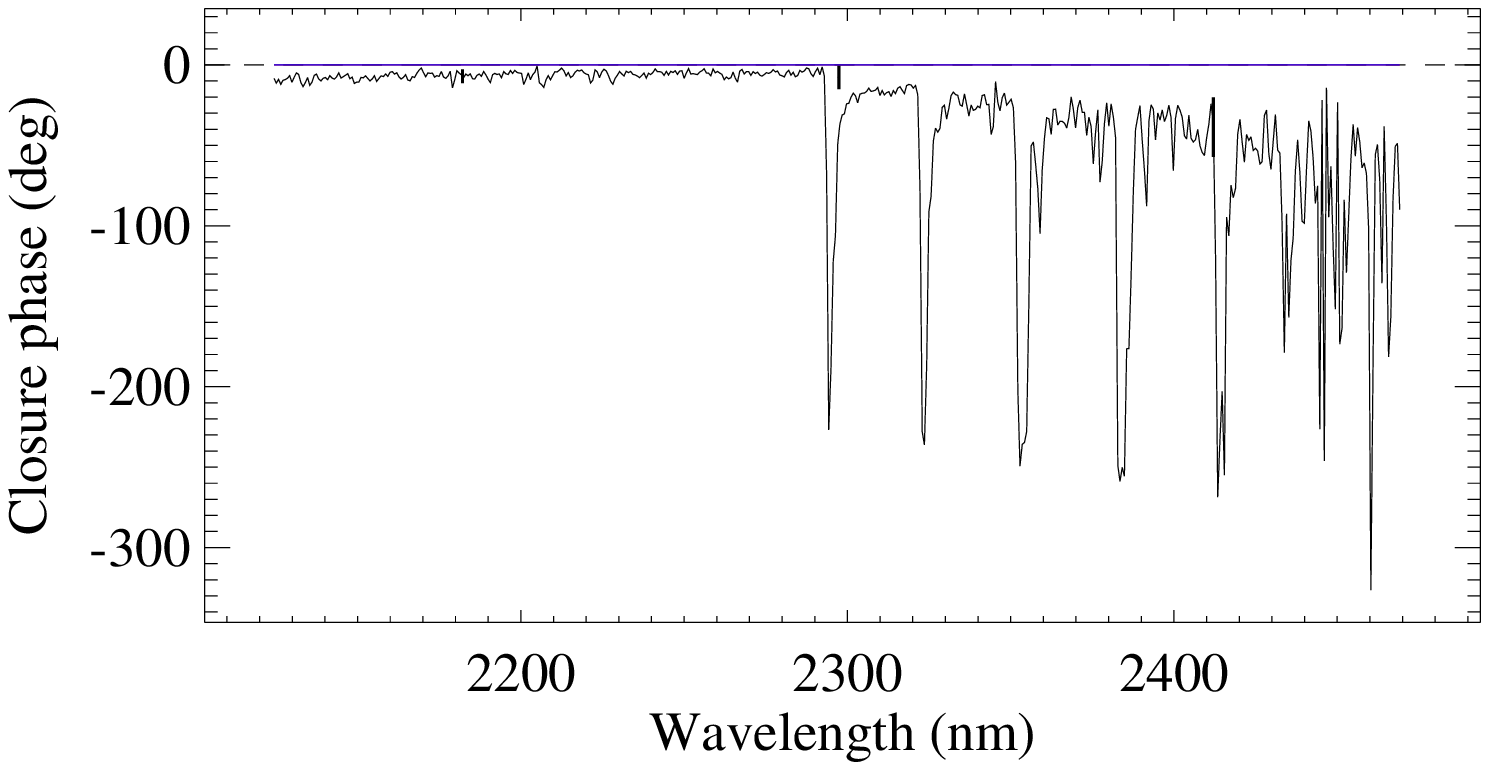}}
\caption{As Fig.~\protect\ref{fig:mr21}, 
but for the MR-K 2.3 $\mu$m mode (PA $-89\deg$).}
\label{fig:mr23}
\end{figure}
}
Figure~\ref{fig:lowres} shows the resulting $H$- and $K$-band squared 
visibility amplitudes 
and closure phases of VY~CMa obtained with the low resolution mode of the 
AMBER instrument. The $J$-band values are not shown because of their lower
quality. Also indicated are the positions of H$_2$O and CO
bands after \citet{lancon00}. The visibility curves show different heights,
corresponding to the different projected baseline lengths. All curves
exhibit a characteristic {\it bumpy} shape that resembles that of recent 
AMBER observations of Mira variables and that are interpreted as being
indicative of the presence of molecular layers lying above the 
continuum-forming photosphere \citep{wittkowski08,wittkowski11}.
Our VY~CMa visibility curves show maxima near 2.25\,$\mu$m and
1.70\,$\mu$m, where the intensity of molecular bands is lowest. The
visibility decreases toward the water bands
centered at $\sim$1.5\,$\mu$m and $\sim$1.9\,$\mu$m. It also exhibits
a steep drop at the onset of the CO absorption feature at 2.3\,$\mu$m.
The decrease of the visibility indicates an increased contribution
from extended intensity at these molecular bands. 
The presence of molecular layers in the extended
atmospheres of red supergiants is consistent with earlier observations
of red supergiants such as $\mu$\,Cep and Betelgeuse 
\citep{perrin04,perrin05,tsuji06} and with the detection of warm water
layers in the ISO and Herschel spectra of VY~CMa \citep{polehampton10,royer10}.

Our VY~CMa closure phases 
exhibit complex curves that correlate with the positions of the
molecular bands and with the projected position angles of the 
observation. Non-0/non-180\,$\deg$ closure phase values are indicative 
of deviations from point symmetry. At the near-continuum band close to 
2.25\,$\mu$m all curves
show relatively small closure phases
with absolute values $<20\deg$. 
This indicates relatively small deviations from point symmetry of 
atmospheric layers close to the photosphere.
These deviations may be caused by 
photospheric convection cells or by 
asymmetric molecular emission that may not be completely
absent at a near-continuum bandpass.
At the molecular bands of H$_2$O and CO, the closure
phase values increase, which indicates a stronger deviation from point
symmetry, and also show variations as a function of position angle.
Along the EW orientation (PA $-90\deg$), the closure phase values
remain at absolute values below $\sim20\deg$ within the CO band 
and the water band.
Toward the NW-SE orientation 
(position angle $-70\deg$), they show a $\sim90\deg$ 
signal at the center of the water band at 1.9\,$\mu$m, and
absolute values up to $\sim$40$\deg$ at the CO band beyond 2.3\,$\mu$m. 
The increased asymmetry
of the molecular layers from the E-W direction toward the NW-SE direction
coincides with the larger-scale asymmetry of the dusty circumstellar envelope
(PA 160\,$\deg$) reported by \citet{wittkowski98} at a wavelength of 2.17\,$\mu$m.
Along the PA of the stronger asymmetry, the sign of the closure
phase differs between the H$_2$O band 
CO band 
indicating that the H$_2$O 
and CO regions are not co-spatial.

Figures~\ref{fig:mr21} and \ref{fig:mr23} (electronic
edition) show the results obtained with the medium resolution modes
at PA $-89\deg$ and $-83\deg$.
These results are consistent
with the low resolution results shown in Fig.~\ref{fig:lowres}.
The visibility shows a maximum at the near-continuum bandpass at
$\sim$2.20--2.25\,$\mu$m and decreases toward the water
band centered at 1.9\,$\mu$m and the CO lines between 2.3\,$\mu$m and
2.5\,$\mu$m, in particular at the CO band-heads.
The corresponding UD curve shows minimum extensions
at 2.20--2.25\,$\mu$m and increases by about 20\% at the
water band near 2.0\,$\mu$m and by up to $\sim$50\% at the CO band-heads.
The closure phase shows relatively small absolute values of 
about 10\,$\deg$ at the near-continuum bandpass, of 30--40\,$\deg$ at the 
water and CO bands, and peaks at the CO band-heads,
confirming asymmetric structures of extended CO band-head emission.
Also shown in Figs.~\ref{fig:mr21} and \ref{fig:mr23}  are visibility fits of 
a UD model and a {\tt PHOENIX} model 
atmosphere
to the 2.20--2.25\,$\mu$m near-continuum band, as described in the
following section.

Overall, our data may qualitatively be consistent with the
scenario proposed by \citet{smith09}, consisting of a cone-like
envelope toward the SE direction that extends from
a large-scale dusty envelope to a small-scale molecular water
and CO envelope close to the photosphere, and additional close
CO cloudlets that may not be co-spatial with the water envelope.
However, for a definite and quantitative picture of the inner
molecular envelope of VY~CMa, an interferometric imaging campaign 
using a well-filled $uv$-plane will be required.
\section{Fundamental parameters}
\label{sec:fundpar}
\begin{figure}
\centering
\resizebox{1\hsize}{!}{\includegraphics{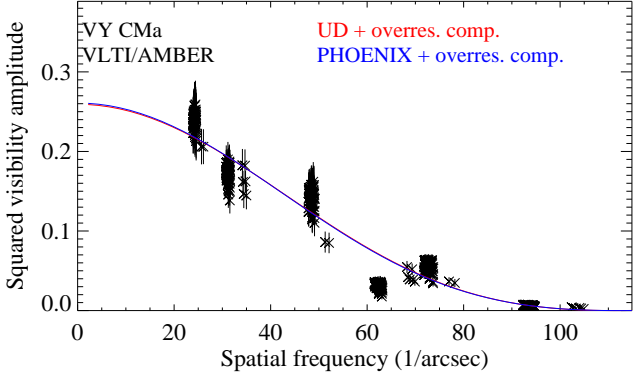}}
\caption{Squared visibility amplitudes of VY~CMa as a function
of spatial frequency obtained at the near-continuum bandpass
at 2.20--2.25\,$\mu$m. 
Also shown are fits of a UD (red) and a {\tt PHOENIX}
atmosphere model (blue). These contribute 50\%
to the total flux, and the remaining part is attributed to an
over-resolved dust component.}
\label{fig:visspfr}
\end{figure}
\onlfig{5}{
\begin{figure}
\centering
\resizebox{1\hsize}{!}{\includegraphics{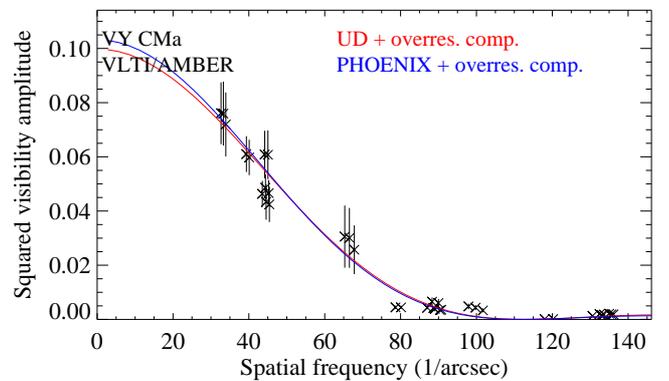}}
\caption{As Fig.~\protect\ref{fig:visspfr}, but for a 
near-continuum bandpass in the $H$-band at 1.70--1.75\,$\mu$m.}
\label{fig:visspfr170}
\end{figure}
}
\begin{figure}
\centering
\resizebox{1\hsize}{!}{\includegraphics{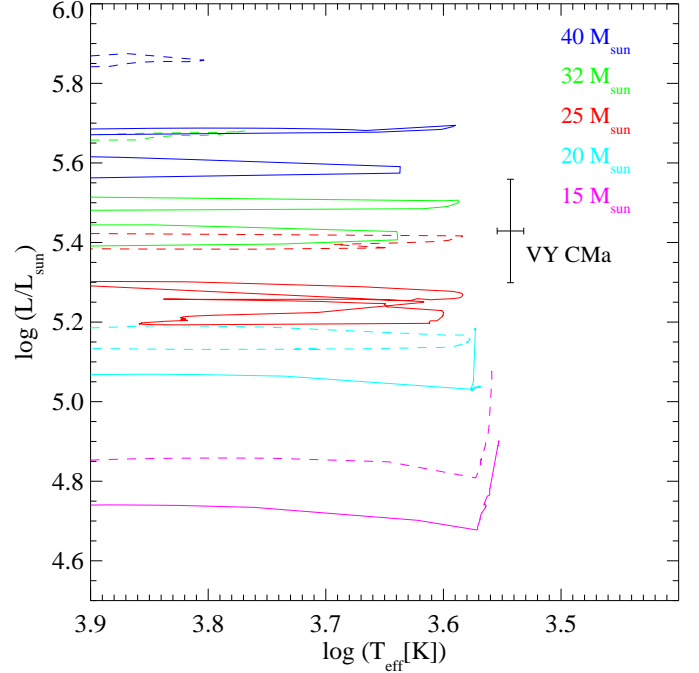}}
\caption{Location of VY~CMa's revised fundamental parameters
in the HR diagram. Also shown are evolutionary tracks from
\citet{ekstrom12} for masses of 15\,$M_\odot$, 20\,$M_\odot$,
25\,$M_\odot$, 32\,$M_\odot$, and 40\,$M_\odot$. The solid
lines denote the models without rotation, and the dashed
lines those with rotation.}
\label{fig:evseq}
\end{figure}
The most direct method to derive the effective temperature is
a measurement of the radius where the Rosseland-mean optical
depth equals $2/3$ together with a measurement of the bolometric
flux. The luminosity is most directly estimated by measurements
of the bolometric flux and the distance. The bolometric flux of
VY~CMa can be well estimated using the high-precision
photometry by \citet{smith01}
complemented by the IRAS fluxes \citep{iras}, cf. also
\citet{jones07} and \citet{choi08}. We de-reddened these data using an
$E_{B-V}$ value of 0.6 obtained from the COBE maps \citep{schlegel98},
and obtained an integrated bolometric flux of 
$f_\mathrm{bol}=(6.3 +/- 0.3)10^{-13}$\,W/cm$^2$.
The distance estimate to VY~CMa has recently 
been improved by high precision parallax measurements by
\citet{choi08} and \citet{zhang12}. 
Here we use the average of these two
measurements, and adopt a distance of $d=1.17_{-0.07}^{+0.08}$\,kpc.
The classic distance used most often in the literature was 
1.5\,kpc going back to \citet{lada78} and the suspected membership to the
open cluster NGC 2362, but whose
distance has also been revised from 1.5\,kpc to 1.2\,kpc
\citep{melnik09}. Our adopted bolometric flux and distance
correspond to a luminosity of $L=(2.7\pm 0.4)10^5 L_\odot$.
This value is slightly smaller than recent estimates
by \citet{choi08} and \citet{mauron11}, mostly because of the 
slightly larger adopted distance taking into account the 
additional measurement by \citet{zhang12}, but otherwise well 
consistent. \citet{choi08}
discussed that this value is also consistent with most earlier
values available in the literature, if they are re-scaled to the
same distance. We note that the error of $L$ is still 
dominated by the uncertainty of the distance.

The angular diameter of VY~CMa was measured by \citet{monnier04} 
in the $K$-band (2.2$\pm$0.3\,$\mu$m) to (18.7$\pm$0.3$\pm$0.4)\,mas.
They note that
the relation of this diameter to a true photospheric diameter relies 
on additional assumption on the limb-darkening, and they also mention
in \citet{massey06} that this value may overestimate the 
photospheric diameter because of a possible contamination by more 
extended molecular layers and dusty features.
Our spectrally resolved visibility data in 
Figs.~\ref{fig:lowres}--\ref{fig:mr23} confirm the presence
of extended layers of H$_2$O and CO, so that a broad-band 
diameter estimate is indeed significantly contaminated by 
extended molecular layers and thus overestimates the photospheric 
diameter. 
With our high spectral resolution, however, we 
are able to estimate a near-continuum diameter at 2.20--2.25\,$\mu$m
that is largely free from contamination by molecular layers.
With our longer baselines (up to 48\,m compared to
38\,m), our data also resolve out possible
compact dusty features to smaller scales and
probe spatial frequencies close to 
the visibility null, which significantly reduces calibration 
uncertainties.
Figure~\ref{fig:visspfr} shows all our visibility data taken at
spectral channels in the near-continuum bandpass between 
2.20\,$\mu$m and 2.25\,$\mu$m as a function of spatial frequency,
together with fits of a UD and a {\tt PHOENIX} model
atmosphere. The {\tt PHOENIX} model was computed with version
16.03.01 of the {\tt PHOENIX} code \citep{phoenix}. 
We used solar metallicity, mass $M=20$\,$M_\odot$, $T_\mathrm{eff}=3500$\,K, 
$\log g=-0.6$\footnote{We computed a large model grid
for different values of $M$, $T_\mathrm{eff}$, and $\log g$,
which will be presented in detail in a forthcoming publication.}.
We used different outer pressure boundary conditions of $10^{-7}$\,dyn/cm$^2$
and $10^{-12}$\,dyn/cm$^2$, giving virtually identical results.
The best-fit diameter using the UD model was 
$\Theta_\mathrm{UD}=(11.0 \pm 0.3)$\,mas, and using the {\tt PHOENIX}
model $\Theta_\mathrm{Ross}=(11.3 \pm 0.3)$\,mas (corresponding to the
model layer where $\tau_\mathrm{Ross}=2/3$). 
Variations of the model mass, $T_\mathrm{eff}$, and $\log g$ lead
to diameter variations of less than 0.1\,mas. 
The flux contribution of the stellar source was a 
free parameter resulting in a value of 51\%. This parameter accounts 
for the fraction of an over-resolved flux component within our FoV 
and also for any residual of the uncorrelated flux outside the FoV
(Sect.~\ref{sec:obs}).
The scatter of the visibility values in Fig.~\ref{fig:visspfr}
may be caused by calibration uncertainties (in particular the points 
at 60--65/arcsec), or by different
contributions of an over-resolved dusty component at different
position angles. However, the diameter remains well constrained by
the measurements close to the visibility null.
Fig.~\ref{fig:visspfr170} (electronic edition) shows the corresponding 
fit using
a near-continuum bandpass in the $H$-band at 1.70--1.75\,$\mu$m, 
resulting in a well consistent angular diameter of 
$\Theta_\mathrm{Ross}=(11.4 \pm 0.3)$\,mas, which increases the
confidence in our diameter estimate.
Figs.~\ref{fig:mr21} and \ref{fig:mr23} also indicate the UD and 
{\tt PHOENIX} models compared to the spectrally resolved 
visibility data. The {\tt PHOENIX} model curve is very close to a 
UD model, and does not predict the large extensions of the 
molecular layers.
{\tt PHOENIX} models predict near-infrared
spectra of red supergiants reasonably well \citep[][and 
Figs.~\ref{fig:mr21} and \ref{fig:mr23}]{lancon07}, which
likely means that opacities from molecular layers are reasonably well
included, but that the extension of the model atmosphere is currently
too compact.  
This effect is most likely caused by 
missing pulsation and/or wind acceleration.
Since our estimate of $\Theta_\mathrm{Ross}$
is based on a near-continuum bandpass, where the contamination
by molecular layers is small, we do not expect this to strongly
affect our estimate of the photospheric diameter. However,
we can not exclude that extended molecular layers
also contribute at our near-continuum bandpass to a small extent,
and that our estimate of $\Theta_\mathrm{Ross}$ may still 
overestimate the photospheric diameter to a small degree (of the order
of 1\,$\sigma$), 
so that it may represent an upper limit of the true photospheric diameter.
The resulting effective temperature, based on $\Theta_\mathrm{Ross}$
and $f_\mathrm{bol}$, is $T_\mathrm{eff}=3490 \pm 90$\,K. 
Using the recent calibration by \citet{vanbelle09}, this value
corresponds to a spectral type of M4, which is consistent
with the spectral classification as M3/M4 by \citet{houk88} as
used in the Simbad database.
If $\Theta_\mathrm{Ross}$ is considered an upper limit, the estimate
of $T_\mathrm{eff}$ is a lower limit.

Figure~\ref{fig:evseq} compares the resulting position of
VY~CMa in the Hertzsprung-Russell (HR) diagram to recent evolutionary sequences
by \citet{ekstrom12}. It is located close to the red limit of recent
evolutionary tracks of initial mass 25\,$M_\odot$ with rotation
(current mass 15\,$M_\odot$) or 32\,$M_\odot$ without 
rotation (current mass 19\,$M_\odot$). 
The position
of VY~CMa is drawn slightly to the right of the Hayashi limit. This 
may be caused by VY~CMa not being in full hydrostatic equilibrium. 
However, a slight remaining underestimate of $T_\mathrm{eff}$,
which we can not fully exclude, could also explain this discrepancy.
\begin{table}
\centering
\caption{Revised fundamental properties of VY~CMa. }
\label{tab:fundpar}
\begin{tabular}{llr}
\hline\hline
Parameter & Value & Reference \\\hline
Distance (kpc) & 1.17 $\pm$ 0.08 & 1 \\
Bolometric flux (10$^{-13}$ W/cm$^2$) & 6.3 $\pm$ 0.3 & 2 \\
Rosseland angular diameter (mas) & 11.3 $\pm$ 0.3& 3 \\
Rosseland radius ($R_\odot$) & 1420 $\pm$ 120 & 4 \\
Effective temperature (K) & 3490 $\pm$ 90& 5 \\
Spectral type & M4 (M3--M4.5) & 6 \\
Luminosity (10$^5 L_\odot$) & 2.7 $\pm$ 0.4& 7\\
Initial mass ($M_\odot$) & 25 $\pm$ 10 & 8\\
Current mass ($M_\odot$) & 17 $\pm$ 8& 9\\
Surface gravity ($\log g$/ cgs) & $-$0.6 $\pm$ 0.4 & 10 \\\hline
\end{tabular}
\tablefoot{1: Average of \citet{choi08} and \citet{zhang12}; 2: Photometry
from \citet{smith01} and IRAS fluxes; 3: This work; 4: 3 with 1;
5: 3 with 2; 6: 5 with \citet{vanbelle09};
7: 2 with 1; 8--10: 5 and 7 with evolutionary tracks 
from \citet{ekstrom12}.}
\end{table}
Table~\ref{tab:fundpar} summarizes our revised fundamental properties
of VY~CMa. With these parameters, VY~CMa would most likely not explode 
at its current stage as a red supergiant, but be at a stage before 
moving blueward in the HR diagram.
\begin{acknowledgements}
This research has made use of the  \texttt{AMBER data reduction package} of the
Jean-Marie Mariotti Center.

\end{acknowledgements}
\end{document}